\documentstyle[pre,aps,floats,graphics,epsfig]{revtex}
\newcommand{\be}{\begin{equation}}
\newcommand{\ee}{\end{equation}}

\author{Matthieu Wyart$^1$, Jean-Philippe Bouchaud$^{1,2}$}
\title{Statistical models for company growth}
\address{
 $^{1}$ Service de Physique de l'\'Etat Condens\'e\\
Orme des Merisiers --- CEA Saclay, 91191 Gif sur Yvette Cedex, France.\\
}
\address{
 $^{2}$ Science \& Finance, Capital Fund Management\\
109--111 rue Victor-Hugo, 92532 Levallois Cedex, France\\
}
\date{\today}
\draft

\begin{document}
\maketitle

\begin{abstract}
We study Sutton's `microcanonical' model for the internal organization of firms, that leads to 
non trivial scaling properties for the statistics of growth rates. We show that the growth rates are
asymptotically Gaussian in this model, at variance with empirical results. We also obtain the 
conditional distribution of the number and size of sub-sectors in this model. We formulate and 
solve an alternative model, based on the assumption that the sector sizes follow a power-law distribution.
We find in this new model both anomalous scaling of the variance of growth rates and non Gaussian
asymptotic distributions. We give some testable predictions of the two models that would 
differentiate them further. We also discuss why the growth rate statistics at the country level and at the company level should be identical. 
\end{abstract}

\section{Introduction}
\label{sect:intro}

The annual growth rate of a company is fluctuating both across companies and from year to
year. It is therefore tempting to study the statistics of this growth rate. It has been
known for many years that the {\it average} growth rate is to a good approximation 
independent of the size of the company. This is known as Gibrat's proportionality law: 
since the growth rate is the {\it relative} size increase of a company (where the size 
refers to the sales, the number of employees, etc.), the fact that the average growth rate
is independent of the size means that on average a company grows proportionally to its size.

A very interesting question, that was only addressed recently, concerns the fluctuations 
of the growth rate, and the size dependence of these fluctuations. Quite remarkably, Stanley et al.
found that the standard deviation $\sigma$ of the growth rate $r$ decreases with the size 
$S$ of the company as $\sigma(S) \sim S^{-\beta}$, with $\beta \approx 0.18$ \cite{Stanley,Stanleybis}. 
This power-law scaling holds 
over six decades, and can be extended to larger sizes by considering countries as 
`companies' and taking the GNP as a measure of the size \cite{Stanley2}. More precisely, the distribution
of the rescaled growth rate $v=r/\sigma(S)$, with $\sigma(S) \sim S^{-\beta}$, 
appears to be size independent. This rescaled distribution $\Pi(v)$ is furthermore found to be
non Gaussian.

This remarkable result is puzzling because one could have naively expected that large companies
(or countries for that matter) would aggregate different independent `shocks' that would 
lead, using the central limit theorem, to a $S^{-1/2}$ decrease of the volatility of 
its growth rate, which would furthermore be Gaussian for large $S$'s. This however assumes 
that a company can be thought of as a collection of $K$ `sub-companies' of average size $S_0$
and weakly correlated activities. In this case, $K=S/S_0$ and if the shocks affecting each
sector of activity have a finite second moment, the central limit theorem applies. 

The fact that $\beta < 1/2$ suggests otherwise. Obviously, if all the sectors of activity 
of a given company had strong cross-correlations, one would find the extreme result that
$\beta=0$. However, this is not the case: Sutton has shown some empirical data that support the
idea that the growths of different sectors are to a good approximation uncorrelated \cite{Sutton}. 
This is what Sutton called the `scaling puzzle', which lead him to propose
a simple model for the internal organization of firms that predicts asymptotically $\beta = 1/4$,
not very far from the empirical result \cite{Sutton}.

The aim of this note is threefold. In a first part which we intend to be also of pedagogical 
interest, we revisit Sutton's model using methods from 
statistical physics, and obtain a number of complementary predictions of this model
that can be compared with empirical data, in particular the distribution of rescaled 
growth rate $\Pi(v)$, which we find to be asymptotically Gaussian, at variance with the empirical result. 
Second, we introduce and study an alternative model where we argue
that the distribution of sizes of the sub-sectors is a power-law, and derive analytically 
the value of $\beta$ and the shape of $\Pi(v)$, which in some regime is found to be
strongly non-Gaussian. We then compare our results to the findings
of Stanley et al. and discuss the plausibility of our alternative model. Finally, we discuss the
interesting fact that GNP growth and company growth behave similarly. This means that the microeconomical
and macroeconomical levels are strongly interconnected. We show that our model is indeed stable upon aggregation.

\section{Sutton's model}
\label{sect:sutton}

We first recall Sutton's model. In the absence of more information, Sutton postulates that 
all partitions of a company of size $S$ in smaller sub-pieces are equiprobable \cite{Sutton}. This is
a kind of `microcanonical', minimum information assumption, similar to the corresponding 
hypothesis in statistical physics where all microstates are equiprobable. For physical systems, this is justified by the Liouville theorem that is itself a consequence of Hamiltonian dynamics; it would be interesting to find an analogue of this theorem for the 
(stochastic) dynamics underlying the organization of firms.

More precisely, Sutton assumes that $S$ is a large integer, and uses known mathematical results on the number of partitions
to compute $\sigma(S)$. Let us show how his results can be recovered directly. For this, we
introduce the following quantity:\footnote{In the following equation, $\delta$
refers to the Dirac delta for continuous variables and to the Kronecker delta
for discrete variables.}
\be
{\cal N}(R,K,S) = \sum_{s_1=1}^\infty \sum_{s_2=s_1}^\infty \cdots \sum_{s_K=s_{K-1}}^\infty 
\delta\left(S - \sum_{i=1}^K s_i\right) 
\int \prod_{i=1}^K P(\eta_i) {\mbox d}\eta_i \,\delta\left(R - \sum_{i=1}^K s_i \eta_i\right).
\ee
This quantity counts the number of partitions of the integer $S$ in exactly $K$ integers
$s_1,s_2,\cdots,s_K$, and such that the total absolute growth rate $R$ is given by the 
sum of independent random variables $\eta_i$ (that we suppose of finite variance), each
weighted by the size $s_i$ of the sub-sector. This assumes a proportionality effect at the
sub-sector level. It will be convenient to introduce the Fourier-Laplace transform of this
quantity (or generating function), defined as:
\be
\hat{\cal N}(q,\nu,\lambda) = \sum_{S=1}^\infty \sum_{K=1}^\infty \int_{-\infty}^{\infty} {\mbox d}R \exp\left[iqR-
\nu K - \lambda S\right] {\cal N}(R,K,S).
\ee
The quantity $\hat{\cal N}(q=0,\nu=0,\lambda)$ is therefore the Laplace transform of the total
number of partitions, and is given by:
\be
\hat{\cal N}(q=0,\nu=0,\lambda) = \sum_{K=1}^\infty \sum_{s_1=1}^\infty \sum_{s_2=s_1}^\infty 
\cdots \sum_{s_K=s_{K-1}}^\infty \exp\left(-\lambda \sum_{i=1}^K s_i\right),
\ee
and be computed explicitly as:
\be
\hat{\cal N}(q=0,\nu=0,\lambda) = \sum_{K=1}^\infty \frac{e^{-\lambda K}}{\prod_{i=1}^K 
(1-e^{-i \lambda K})}.
\ee
For $\lambda \to 0$, the sum over $K$ can be approximated by an integral:
\be\label{1}
\hat{\cal N}(q=0,\mu=0,\lambda) \approx \int_{0}^\infty {\mbox d}K \exp\left(-\lambda K
- \int_0^K  {\mbox d}x \ln (1-e^{-\lambda x})\right).
\ee
Now the integral over $K$ can be estimated using a saddle-point 
method. The saddle point $K^*$ obeys the following equation:
\be
\lambda = - \ln (1-e^{-\lambda K^*}),
\ee
which for small $\lambda$ gives: 
\be
K^* \approx \frac{1}{\lambda} \ln \frac{1}{\lambda}.
\ee
Plugging this result in Eq. (\ref{1}) leads to:
\be
\hat{\cal N}(q=0,\nu=0,\lambda \to 0) \sim \exp\left(\frac{1}{\lambda}\int_0^\infty 
{\mbox d}v \ln (1-e^{-v})\right)
= \exp\left(\frac{\pi^2}{6\lambda}\right),
\ee
where we have neglected preexponential corrections, that can also be computed. Now, it is
easy to check that the inverse Laplace transform of $\hat{\cal N}(q=0,\nu=0,\lambda \to 0)$
behaves, for large $S$, as:
\be
{\cal N}(S) \sim \exp[b \sqrt{S}], \qquad b = \pi \sqrt{\frac23}
\ee
which is the Hardy-Ramanujan result at large $S$ \cite{partition}. In the course of the
calculation, one also discovers that, as far as scaling is concerned, $\lambda \sim S^{-1/2}$.
One could extend the computation to get the exact prefactor, equal to $(4 \sqrt{3} S)^{-1}$. 

One can easily extend the computation to $\nu \neq 0$. The saddle point is now at:
\be
K^* \approx \frac{1}{\lambda} \ln \frac{1}{\lambda+\nu}.
\ee
Now, setting $\nu = x \lambda/|\ln \lambda|$, one finds, in the limit $\lambda \to 0$, 
\be\label{2}
\frac{\hat{\cal N}(q=0,x \lambda/|\ln \lambda|,\lambda)}
{\hat{\cal N}(q=0,0,\lambda)} \approx e^{-x}.
\ee
Having noted that $e^{-x}$ is the Laplace transform of $\delta(u-1)$, we 
conclude that when $S \to \infty$, the variable $K/\sqrt{S} \ln S$ tends to unity
with probability one. One can also study how the fluctuations behave for large $S$.
Setting $K = \sqrt{S} \ln S + y \sqrt{S}$, one finds that the Laplace transform $\hat P(z)$ of the
distribution $P(y)$ of the random variable $y$ reads, for $S \to \infty$:
\be
\hat P(z)=\int {\mbox d}y e^{-zy} P(y) = \exp\left[-z + (1+z) \ln (1+z)\right],
\ee
which shows that the distribution of $y$ is non Gaussian, even in the limit $S \to \infty$.
For example, the skewness of $P(y)$ is found to be equal to $-1$. 
In summary, we find that the average number of `sub-entities' is equal to $\sqrt{S}\ln S$, 
with relative (non Gaussian) fluctuations which go to zero as $1/\ln S$. 
The average size of a sub-piece is clearly equal to $\sqrt{S}/\ln S$. 

Therefore, the most probable 
partition of a large 
integer $S$ is to break it in $\sqrt{S}$ parts of size $\sqrt{S}$ (neglecting logarithms) \cite{branchedpolymer}. In fact, as we now show, this is not really correct. A better description is to say that one has
$\sqrt{S}$ pieces of size $1$, $\sqrt{S}/2$ pieces of size $2$, ... and one piece of size $\sqrt{S}$.
More precisely, what is the average number of occurrences $N(s|S)$ of a piece of size $s$, 
given the total size $S$? A little reflection tells us that this is given by:
\be
N(s|S) \equiv {\cal N}(S) \langle O(s|S) \rangle  = 
\sum_{k=1}^\infty {\cal N}(S-ks) \equiv \sum_{k=1}^\infty k Q(S-ks),
\ee 
where $Q$ is the probability of occurrence that the number $s$ appears exactly $k$ times in 
the partition, defined by the above equation. For $1 \ll s \ll S$, $N(s|S)$ can be approximated by:
\be
N(s|S) \approx {\cal N}(S) \sum_{k=1}^\infty \exp(-\frac{bks}{2\sqrt{S}}) 
\sim \frac{{\cal N}(S)}{\exp(\frac{bs}{2\sqrt{S}})-1}.
\ee
One therefore finds the following interesting result: the size distribution of 
sub-sectors follows, in Sutton's model, a Bose-Einstein distribution. This
distribution behaves as a power law $1/s$, for $s \ll \sqrt{S}$ and decays
exponentially fast for $s \gg \sqrt{S}$. This is a directly testable prediction
of Sutton's model. One can furthermore check directly that:
\be
\frac{\sum_{s=1}^S s N(s|S)}{{\cal N}(S)} \approx \frac{4S}{b^2} 
\int_0^\infty {\mbox d}u \frac{u}{e^u-1} = S,
\ee
as it should. 

As noticed by Sutton, the quantity $N(s|S)$ is interesting because it allows us to compute the variance $\sigma_R^2(S)$ of the absolute growth rate $R$, defined as:
\be
\sigma_R^2(S)=\overline{\langle R^2 |S \rangle},
\ee
where the brackets means an average over the random growth rates $\eta_i$ and 
the overline is an average over all partitions of $S$. Using the fact that the 
$\eta_i$'s are independent and of variance equal to $\sigma_0^2$, one has:
\be
\overline{\langle R^2 |S \rangle}  = \sigma_0^2  \sum_s s^2 \frac{N(s|S)}{{\cal N}(S)} 
\approx \sigma_0^2 \sum_s \frac{s^2}{\exp(\frac{bs}{2\sqrt{S}})-1} = \frac{2^{5/2}3^{3/2} \zeta(3)}
{\pi^3} \, \sigma_0^2 S^{3/2}=  1.13955....\, \sigma_0^2 S^{3/2}.
\ee
This is Sutton's result: the conditional variance of the absolute return grows as $S^{3/2}$,
therefore the variance of the {\it relative} return $r=R/S$ decays as $S^{-1/2}$, which 
is equivalent to the statement that $\beta=1/4$ \cite{Sutton}. In intuitive terms, the total absolute 
return is the random sum of $\sim \sqrt{S}$ different terms, all of order $\sim \sqrt{S}$,
which gives a random number of order $\sqrt{S} \sqrt{\sqrt{S}} \sim S^{3/4}$. This rough (and
slightly incorrect) argument actually suggests that the absolute growth rate is the sum of a large number ($\sqrt{S}$) of
random variables, and therefore should be Gaussian for large $S$. We can show 
this more precisely by computing the kurtosis $\kappa$ of $R$, defined as:
\be
\kappa = \frac{\overline{\langle R^4 |S \rangle}}{\overline{\langle R^2 |S \rangle}^2}-3.
\ee
We assume that the individual growth rates $\eta$ have a finite kurtosis given by
$\kappa_0$. Therefore:
\be
\overline{\langle R^4 |S \rangle} = \sigma_0^4 \left[ (3+\kappa_0) 
\overline{\sum_{i} s_i^4} + 3 \overline{\sum_{i \neq j} s_i^2 s_j^2} \right]= 
\sigma_0^4 \left[\kappa_0 \overline{\sum_{i} s_i^4} + 3 
\overline{\sum_{i,j} s_i^2 s_j^2} \right].
\ee  
The first term is easy to compute using $N(s|S)$ and one finds:
\be
\overline{\sum_{i} s_i^4} = \sum_s s^4 \frac{N(s|S)}{{\cal N}(S)} \approx 
\frac{2^{11/2} 3^{7/2}}{\pi^5} \zeta(5) \, S^{5/2} = 7.17114....\, S^{5/2}
\ee
The second term is more subtle since one needs to know the correlation of 
the number of occurrences of two integers $s,s'$ involved in the partition of $S$,
$\langle O(s|S) O(s'|S) \rangle$.
This quantity can be obtained similarly to $N(s|S)$. For $s=s'$, one has:
\be
{\cal N}(S) \langle O(s|S)^2 \rangle \equiv \sum_{k=1}^\infty k^2 Q(S-ks) = 
\sum_{k=1}^\infty (2 k -1) {\cal N}(S-ks).
\ee
which can be again be approximated as
\be
\langle O(s|S)^2 \rangle \approx 
\sum_{k=1}^\infty (2k-1) \exp\left(-\frac{bks}{2\sqrt{S}}\right)=
\frac{\exp(\frac{bs}{2\sqrt{S}})+1}{\left(\exp(\frac{bs}{2\sqrt{S}})-1\right)^2}
\ee
Therefore:
\be
\sum_{s=1}^S s^4 \left[\langle O(s|S)^2 \rangle - \langle O(s|S) \rangle^2 \right]\approx
\left(\frac{2 \sqrt{S}}{b}\right)^5 \int_0^\infty {\mbox d}u u^4
\frac{e^u}{(e^u-1)^2} = \frac{2^{9/2} 3^{3/2}}{5 \pi}  \, S^{5/2} = 7.48509.... \, S^{5/2}.
\ee
The  terms with $s \neq s'$ can be computed from:
\be
\langle O(s|S) O(s'|S) \rangle \approx \sum_{k=1,\ell}^\infty \exp\left(-\frac{bks}{2\sqrt{S}}
-\frac{b\ell s'}{2\sqrt{S}} - \frac{b}{8 S^{3/2}} (ks + \ell s')^2 + ...\right).
\ee
The reason we took one extra term in the above expansion is that 
$\langle O(s|S) O(s'|S) \rangle - \langle O(s|S) \rangle \langle O(s'|S) \rangle$
is zero to first order. The non zero correlation comes from the term
$k \ell s s'$ in the above expression. To lowest order, one finally finds,
\be
\sum_{s \neq s'=1}^S s^2 s'^2 \left[\langle O(s|S) O(s'|S) \rangle - 
\langle O(s|S) \rangle \langle O(s'|S) \rangle \right] \approx 
- \frac{2^{19/2} 3^{5/2}}{\pi^5} [\zeta(3)]^2 S^{3/2} = - 53.2953...S^{3/2}.
\ee
This contribution is a factor $S$ smaller than the other two contributions, but
has a rather large prefactor. To leading order in $S$, the final result reads:
\be
\kappa \approx \frac{1}{\sqrt{S}}  \left[5.52232...\kappa_0 + 5.76408...\right]
\ee
The conclusion is that if the growth rates of the sub-sectors are non Gaussian, the
kurtosis of the aggregate growth rate decreases as $1/\sqrt{S}$ for large $S$. This
is expected on general grounds, since we have seen above that the number of 
independent sub-entities is of order $\sqrt{S}$, and the kurtosis of a sum 
decreases as the inverse of the number of independent terms in a sum. The second
contribution comes from the fluctuations of the numbers of terms in the sum.

Therefore, {\it asymptotically}, the rescaled aggregate growth rate $rS^{1/4}$ is found to be 
Gaussian in Sutton's model, at variance with empirical findings. However, for finite $S$, there
are important corrections to this asymptotic result: suppose that the initial 
kurtosis of $\eta$ is equal to $3$, which is the case when  $\eta$ is
distributed according to a symmetric exponential. Take a reasonable value $S = 100$.
Then the residual kurtosis of the growth rate is still quite large, $\sim 2.2$. Hence,
significant deviations from a Gaussian distribution may be observed in reality, but should 
diminish as $S$ becomes large. We shall
come back to this issue in section ~\ref{sect:discussion}.

\section{An alternative model}
\label{sect:levy}

\subsection{Definition of the model}

We now discuss another model, where we assume that companies are formed by aggregating 
entities that have a certain a priori distribution of sizes, that we choose to be a power-law. 
The motivation for this is two-fold. First, the distribution of company sizes in a country 
is known to be a Pareto (power-law) distribution. Since the scaling law for the variance of the
growth rate also seems to hold at the country level, one could indeed argue that the actual distribution
of company sizes should play a role. Second, there is a quite general and plausible {\it dynamical} model 
that leads to a power-law distribution of sizes. Assume, as in Sutton's model, that each sub-entity in 
a company has a random growth rate. The role of the business management is, to a certain extent, to redistribute
the income of each sector of activity such as to help the less performing ones to catch up. Therefore, 
a reasonable dynamical model for the size $s_i(t)$ of a given sub-entity is:
\be
\frac{{\mbox d}s_i}{{\mbox d} t} =  \gamma \left({\frac{1}{K}}\sum_{j=1}^K s_j(t) - s_i(t)\right) + \eta_i(t) s_i(t),
\ee
where the first two terms describe redistribution of resources among the sub-entities, and the last
term the random growth rate. The parameter $\gamma$ measures the strength of the redistribution policy. 
It can be shown that the stationary distribution for such a stochastic process has a power law tail, 
$p(s) \sim s^{-1-\mu}$, with $\mu=1+\gamma/\sigma_0^2$. (See the detailed discussion and generalization in
 \cite{BM2000}, and also \cite{Gabaix,QF} for alternative models.)

Hence, we assume that the a priori distribution of the size of sub-entities has a power law tail:
\be
p(s) \approx \frac{\mu s_0^\mu}{s^{1+\mu}} \qquad (s \to \infty), 
\ee 
We also assume that a company is composed of an arbitrary number $K$ of such sub-entities, with a certain a priori weight ${\cal Q}(K)$. This means that if one chooses randomly a company in a country, there is a probability proportional to ${\cal Q}(K)$ for this company to contain exactly $K$ sectors. We will see below that ${\cal Q}(K)$ can be inferred from empirical data. The unnormalized distribution of growth rates 
for a given company size $S$ reads, in this new model:
\be\label{defmodel}
{\cal N}(R,S) = \sum_{K=1}^\infty {\cal Q}(K)  \int \prod_{i=1}^K p(s_i){\mbox d}s_i
\,\delta\left(S - \sum_{i=1}^K s_i\right) 
\int \prod_{i=1}^K P(\eta_i) {\mbox d}\eta_i\, \delta\left(R - \sum_{i=1}^K s_i \eta_i\right),
\ee

\subsection{Distribution of company sizes}

Let us first establish some results on the size distribution of companies ${\cal N}(S)=\int {\mbox d}R {\cal N}(R,S)$, a quantity much studied in a different context in \cite{Bardou}. This will also enable us to 
relate ${\cal Q}(K)$ to this empirically observable quantity.
We first study the case the simplest case  ${\cal Q}(K)$=1.
The following results are obtained using Laplace transforms, as above.
We write:
\be
\hat{\cal N}(q,\lambda) = \int_0^\infty {\mbox d}S \int_{-\infty}^{\infty} {\mbox d}R \exp\left[iqR- \lambda S\right] {\cal N}(R,S) = \sum_{K=1}^\infty \left[\int {\mbox d}s {\mbox d}\eta p(s) P(\eta) \, e^{iq s \eta - \lambda s}\right]^K.
\ee
For simplicity, we assume that $P(\eta)$ is Gaussian with unit variance, and introduce the quantity $g(q,\lambda)$ as:
\be
g(q,\lambda) =  \int {\mbox d}s \, p(s)  \left[1-e^{-q^2 s^2/2 - \lambda s}\right],
\ee
in terms of which one finally has:
\be
\hat{\cal N}(q,\lambda) = \frac{1}{g(q,\lambda)}.
\ee
All the following asymptotic results will only depend on the behaviour of $g(q,\lambda)$ in the limit $q, \lambda \to 0$. If we first study the case $q=0$ from which ${\cal N}(S)$ is deduced, 
one finds that one has to distinguish the cases $\mu < 1$ and $\mu > 1$ \cite{Bardou}. The small $\lambda$ behaviour of 
$g$ is found to be:
\be
g(q=0,\lambda) \approx  \Gamma(1-\mu) (s_0\lambda)^\mu \quad (\mu < 1); \qquad 
g(q=0,\lambda) \approx  \lambda \langle s \rangle \quad (\mu > 1).
\ee
For $\mu > 1$, the average size of a sub-entity is finite and equal to $\langle s \rangle$. 
Inverting the Laplace transform then leads to:
\be
{\cal N}(S) \approx \frac{1}{\langle s \rangle}, 
\ee
for $\mu > 1$, whereas for $\mu < 1$, one has:
\be
{\cal N}(S) \approx \frac{\sin \pi \mu}{\pi} \frac{1}{s_0} \left(\frac{s_0}{S}\right)^{1-\mu}.
\ee
The case $\mu=1$ is special and involves logarithmic corrections. Intuitively, the difference of behaviour
comes from the fact that when $\mu > 1$, the typical number of sub-entities behaves as $K \sim S/\langle s \rangle$, whereas when $\mu < 1$, a single sub-entity represents a sizeable fraction of the whole company and $K \sim S^\mu \ll S$.

Assuming now that ${\cal Q}(K)$ decays as a power-law ${\cal Q}(K) \sim K^{-1-\alpha}$ and that $\mu > 1$, 
we find, using the same method, the following result for $P(S)$: 
\be\label{cond-dist}
{\cal N}(S) \sim \frac{1}{S^{1+\alpha}} \quad (\alpha \leq \mu); \qquad
{\cal N}(S) \sim \frac{1}{S^{1+\mu}} \quad (\alpha \geq \mu).
\ee 
The case $\alpha \geq \mu$ corresponds to a situation where large companies only contain a small
number of sectors (see below). This is not very plausible; furthermore, this would lead to a variance of the 
growth rate $R$ that grows proportionally to size $S$, i.e. $\beta=0$, which is not 
compatible with empirical data. Therefore we will assume in the following $\alpha \leq \mu$. In this case there is a direct relation between the tail of ${\cal Q}(K)$ and the tail of the size distribution of companies. 
Empirically, $\alpha$ is found to be close to unity: $\alpha \approx 1.05$ \cite{Axtell}.

\subsection{Fluctuations of the growth rate}

We now turn to the prediction of this model for the growth rate fluctuations. One need to consider 
three cases: $\mu>2$, $1 <\mu < 2$, $\mu<1$. The case $1 <\mu < 2$ is, as we will show, the interesting one. In the relevant situation where $\alpha \leq \mu$, one can show that the value of $\beta$ and the shape of the rescaled distributions  are independent of the value of $\alpha$, and we choose in the following, for simplicity ${\cal Q}(K)$ to be constant.

We now need to study $\hat{\cal N}(q,\lambda)$ with $q \neq 0$, that gives access to the distribution of the growth rate. This case can be treated by 
identifying the correct scaling region in the $q,\lambda$ plane, which means, in concrete terms, 
the scaling relation between $R$ and $S$. For example, when $\mu > 2$, one expects the Central Limit Theorem
to hold, suggesting $R \sim \sqrt{S}$. So we set $q = \theta \sqrt{\lambda}$, and take the limit 
$\lambda \to 0$. If $\mu > 2$, one finds:
\be
g(q,\lambda) \approx \lambda \left[\langle s \rangle + \frac{\theta^2}{2} \langle s^2\rangle \right] + \cdots
\ee
where the $\cdots$ refers to higher order terms in $\lambda$, the precise form of which depend on the
value of $\mu$. Therefore, in this regime, 
\be\label{eqG1}
\hat{\cal N}(q,\lambda) \approx \frac{1}{\lambda \left[\langle s \rangle + \frac{\theta^2}{2} \langle s^2\rangle \right]}.
\ee
Now, we introduce the probability $P(R|S)$ to observe a certain growth $R$ given $S$. Then, by definition,
\be
\hat{\cal N}(q,\lambda) = \int_0^\infty {\mbox d}S \int_{-\infty}^{\infty} {\mbox d}R \exp\left[iqR- \lambda S\right] {\cal N}(S) P(R|S).
\ee
Assuming that $P(R|S)=S^{-1/2} \Pi(R S^{-1/2})$ and using the above result for ${\cal N}(S)$ leads to:
\be\label{eqG2}
\hat{\cal N}(q=\theta \sqrt{\lambda},\lambda) \approx  \frac{1}{\lambda \langle s \rangle} 
\int_0^\infty {\mbox d}u \int_{-\infty}^\infty {\mbox d}v \exp\left[i\theta v \sqrt{u}- u\right] \Pi(v),
\ee
where we have set $\lambda S=u$ and $R = v \sqrt{S}$. It is now easy to see that Eqs.~(\ref{eqG1}), ~(\ref{eqG2}) are satisfied if:
\be
\Pi(v) = \frac{1}{\sqrt{2 \pi \sigma^2}}\exp(-\frac{v^2}{2 \sigma^2}) \qquad \sigma^2 = \frac{\langle s^2\rangle}
{\langle s\rangle}.
\ee
Therefore, in the case $\mu > 2$, the variance of the relative growth rate decreases as $S^{-1/2}$ (i.e. 
$\beta=1/2$), and the distribution of growth rates is Gaussian.

More interesting is the case $1 < \mu < 2$. It turns out that in this regime, the correct scaling is 
$P(R|S)=S^{-1/\mu} \Pi(R S^{-1/\mu})$ and $q = \theta \lambda^{1/\mu}$. In this regime, one now has, for 
small $\lambda$:
\be\label{eqL1}
g(q,\lambda) \approx \lambda \left[\langle s \rangle + \mu {|s_0\theta|^\mu} I(\mu) \right] + \cdots,
\ee 
where
\be
I(\mu) = \int_0^\infty \frac{{\mbox d}t}{t^{1+\mu}} (1-e^{-t^2/2})=-2^{-1-\mu/2} \Gamma(-\frac{\mu}{2}).
\ee
Now, it is easy to show that the scaling function $\Pi(v)$ is now precisely a symmetric Levy stable distribution of index $\mu$, $L_\mu(v)$. This comes from the fact that the Fourier transform of $L_\mu(v)$  gives $\exp(-A u |\theta|^\mu)$, where $A$ is a constant, so that the integral over $u$ in Eq.~(\ref{eqG2}) now reproduces Eq.~(\ref{eqL1}). 

However, this is not the whole story. The reason is that a direct computation of the variance of $R$ 
(from the derivative of $g(q,\lambda)$ with respect to $q^2$ at $q=0$) leads an apparently 
contradictory scaling, since:
\be\label{scalingmu}
\langle R^2 | S \rangle  \propto S^{3-\mu}
\ee
instead of $S^{2/\mu}$ as one might have naively expected from the scaling form of $P(R|S)$. One
should now remember that L\'evy stable distributions $L_\mu(v)$ with $\mu < 2$ have tails decaying as $v^{-1-\mu}$, and
thus a formally infinite variance. This means that $\langle R^2 | S \rangle$ is actually dominated by the region where $R$ is of order $S$, such that indeed:
\be
\langle R^2 | S \rangle \approx S \int_{-S}^{+S} \frac{R^2 {\mbox d} R}{|R|^{1+\mu}} \sim S^{3-\mu}.
\ee
Therefore, the L\'evy stable distribution only holds in the scaling region $R \sim S^{1/\mu}$. 
For $R \sim S \gg S^{1/\mu}$ the distribution is truncated. When $\mu \to 1$, the truncation 
`invades' the scaling regime, and the result becomes again different for $\mu < 1$, see below. 

The conclusion of this analysis is that the variance of the relative growth rate $r=R/S$ scales 
in this regime 
with an exponent $\beta=(\mu-1)/2$, that interpolates between the standard value $\beta=1/2$ for $\mu=2$ 
and $\beta=0$ for $\mu =1$ (although these marginal cases are affected by logarithmic corrections). 
However, the surprising result is that in this regime the distribution of $R$ does {\it not} re-scale 
as a function of $rS^\beta$ but rather as $rS^{(\mu-1)/\mu}$.\footnote{For other situations where this `anomalous scaling' occur, see \cite{PhysRep,GL}.} We will discuss this in relation with
empirical results in the next section. 

When $\mu < 1$, it is easy to show that now $R \sim S$, i.e. $\beta=0$, which disagrees with empirical 
results. Furthermore, the result one finds for the
scaling function $\Pi$ is no longer universal. When $\mu > 1$, the scaling function was universal 
in the sense that its shape only relied on the finiteness of the variance of $\eta$. When $\mu < 1$, on
the other hand, only a finite number of terms (sub-entities) contribute to the sum $R$, and one
cannot expect a Central Limit Theorem to hold. When $P(\eta)$ is Gaussian of variance $\sigma_0^2$, all
moments of $P(R|S)$ can be computed using the method of \cite{Derrida}. One finds for example:
\be
\langle R^{2} | S \rangle = \frac{1-\mu}{1+\mu} \sigma_0^2 S^{2} 
\ee 
and:
\be
\langle R^{4} | S \rangle =  3 \left[\frac{(3-\mu)(2-\mu)(1-\mu) + 2\mu(1-\mu)^2}{(3+\mu)(2+\mu)(1+\mu)}
\right]  \sigma_0^4 S^{4} 
\ee
Numerically, we have found that $P(R|S)$ could be rather well fitted by a `stretched Gaussian' form,
$\exp-(R/S)^\alpha$ with $\alpha < 2$. For example, for $\mu=1/2$, we found $\alpha \approx 4/3$.
This cannot be exact, however, since the exact kurtosis is found to be $1.37143$, whereas the kurtosis 
of the stretched Gaussian with $\alpha=4/3$ is $1.22219$. Note that $\langle R^{4} | S \rangle$ would have a different value if $P(\eta)$ was non Gaussian: this shows that 
for $\mu < 1$ the distribution $\Pi(v)$ is non universal.

\subsection{Conditional distribution of sector sizes}

Finally, one can also compute in this model the conditional distribution of sector sizes, $P(s|S)$, which depends on the value of $\alpha$. When $\alpha \leq \mu$, we find that $P(s|S)$ 
is the sum of two contributions: one power-law regime $s^{-1-\mu}$ for $s \ll S$ which reflects the a priori distribution of sector sizes, and a small `hump' for $s \sim S$ of height which vanishes for large $S$:
\be\label{pss}
P(s|S) \approx \frac{\mu s_0^\mu}{s^{1+\mu}}, \quad (s \ll S); 
\qquad P(s|S) \approx \frac{F(s/S)}{S^{1+\mu-\alpha}}\quad (s \sim S),
\ee
where $F(.)$ is a certain scaling function of order unity, that vanishes for
$s > S$. For $\mu  < \alpha$, one the other hand, the hump survives when $S \to \infty$ whereas 
the power-law regime disappears.
In other words, when $\alpha$ is larger than $\mu$, the typical number of sectors $K^*$ tends to be small 
and the typical size of the sectors is of the order of $S$ itself.

\subsection{Stability upon aggregation}

As mentioned in the introduction, the scaling of GNP growth rates is empirically found to be
very similar to the scaling of company growth \cite{Stanley2}. In this respect, it is worth noting that Sutton's construction
is not stable upon aggregation: aggregating companies characterized by an exponent $\beta=1/4$ using Sutton's prescription at the country level, leads to an exponent $\beta=3/8$. In our model, on the other hand, stability upon aggregation is by construction satisfied. The argument is very simple, and relies on the fact that the
results are independent of the value of the company size exponent $\alpha$, provided $\alpha < \mu$. The idea is to consider the GNP itself as the sum of independent sectors, i.e. to remove the `shells' that define 
companies, which are an intermediate level of clustering. A country is therefore in this description a `super-company' with many sectors. The sectors are the same than previously, so they have the very same Pareto tail of exponent $\mu$ for their size distribution. Now we just have to assume that there is a given distribution $Q'(K)$ that describes the distribution of the number of independent sectors in different countries. If $Q'(K)$ has a Pareto tail with exponent $\alpha'$ with $\alpha ' < \mu$, we can repeat the above arguments and find the 
same value for the exponent $\beta=(\mu-1)/2$ at the country level.

\section{Discussion -- Comparison with empirical data}

\label{sect:discussion}

We have shown how several interesting asymptotic predictions of Sutton's 
model could be 
derived. Apart from Sutton's central result, namely that the root mean square 
of the growth
rate decreases with the company size $S$ as $S^{-1/4}$ (i.e. $\beta=1/4$), we 
have shown that
the distribution of growth rate should be asymptotically Gaussian, with a 
kurtosis
that decays as $S^{-1/2}$. The first result is, as noticed by Sutton, in 
rather good agreement 
with the empirical results of Stanley et al.\cite{Stanley}, although the value of the exponent $\beta$ 
is closer to $0.18$. The second result is however problematic, since in this 
model one should
find a rescaled distribution of growth rates that progressively deforms with 
$S$ 
as to become Gaussian for very large $S$, whereas the data indicates that the 
rescaled 
distribution is actually to a good approximation independent of $S$ {\it and} 
non Gaussian. A closer look at the data of Stanley et al. in fact suggests that non 
Gaussian tails are {\it more pronounced} for larger companies \cite{Stanleybis}.

We have then explored an alternative to Sutton's model, where the size of the 
`sub-entities' 
is postulated to be a power-law (Pareto) with an exponent $\mu$. This is 
motivated by the 
ubiquitous observation of Pareto distributions for company sizes, and by a 
simple dynamical
model that indeed leads to a stationary power-law distribution of sizes. In 
this respect, it
is not obvious how one would write a natural dynamics for sector growth that 
leads to 
Sutton's `microcanonical' ensemble where all partitions are equiprobable. As 
a function of
$\mu$, we have found three qualitatively different regimes. In particular, 
when $1 \leq \mu \leq 2$, we
find that $\beta=(\mu-1)/2$. The empirical value $\beta=0.18$ corresponds to 
$\mu=1.36$, which is indeed larger than the value of $\alpha \approx 1.05$ reported for firm sizes in \cite{Axtell}, as required for the consistency of our analysis. Our 
model then predicts an $S$ independent distribution for the growth rate 
multiplied by $S^{(\mu-1)/\mu}$
(and not by $S^\beta$), which is a symmetric L\'evy stable distribution. Note 
however that 
these are asymptotic results that require $S^{(\mu-1)/\mu} \gg 1$, such that 
the scaling 
region is not affected by truncation effects (see the discussion after Eq.~\ref{scalingmu}). 
For finite $S$ and $\mu$ close to one, one expects strong finite size effects, and a very slow convergence
towards the asymptotic value. This is why numerical simulations are needed to explore the moderate $S$
regime. We show in Fig. 1 the distribution of rescaled 
returns obtained from a numerical simulation for $\mu=1.3$, a Gaussian $P(\eta)$ and 
for $S \sim 500$ and $S=5000$. Notice $\Pi(v)$ can be very roughly 
approximated by a symmetric exponential for small enough $S$: $\Pi(v)=\exp(-|v|/v_0)$, as suggested by the empirical data. 
The systematic deviations from this 
form both at small values of $v$ and at large $v$ are qualitatively similar to the ones 
observed empirically (see \cite{Stanleybis}): $\Pi(v)$ 
is actually parabolic for small values of $v$ and decays slower than 
exponentially at large $v$. One also observes strong finite size effects: as $S$ increases, the tail of 
the distribution becomes fatter and fatter. This is expected since asymptotically this distribution
should converge to a L\'evy distribution with a power-law tail, which is indeed fatter than an
exponential. Note that we expect only qualitative agreement with empirical data, since the assumption 
that $P(\eta)$ is Gaussian at the sector level is probably incorrect and does influence the detailed shape of $\Pi(v)$ for
finite $S$. However, as mentioned above, the systematic `fattening' of the tails as $S$ becomes larger
seems to be present in the empirical results of Stanley et al. \cite{Stanleybis}.

\begin{figure}
\begin{center}
\psfig{file=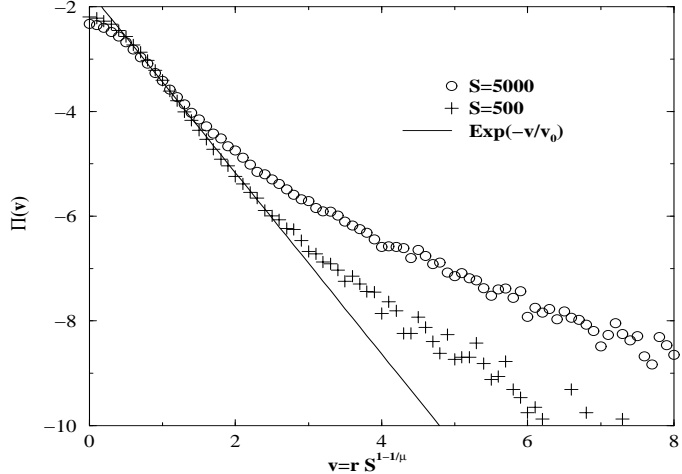,width=9.cm,height=6.5cm} 
\end{center}
\caption{Distribution $\Pi(v)$ of the rescaled returns for $v>0$ and two values of $S$: $S \approx 
500$ and $S \approx 5000$, and for $\mu=1.3$. A simple exponential form, as suggested in \protect\cite{Stanley}, is shown for comparison. Note that for this range of $S$, the distributions do not re-scale, except in the `central' region. For large values of $S$, the distribution should converge towards a L\'evy distribution of index $\mu=1.3$: one can clearly see the tails getting fatter as $S$ increases.}
\label{Fig1}
\end{figure}

It would be extremely interesting to obtain direct empirical information on 
the conditional 
distribution of the size $s$ and total number $K$ of the sub-entities for a fixed $S$. We 
have seen that Sutton's 
model predicts a Bose-Einstein distribution for $s$, that behaves as $1/s$ 
for $s \ll \sqrt{S}$,
and beyond which it falls rapidly, whereas $K$ becomes peaked around the value $\sqrt{S} \ln S$. 
In our model, on the other hand, the  conditional 
distribution of $s$ is, as soon as $\mu \geq \alpha$ and for $s \ll S$, identical to 
the a priori distribution $p(s) \sim s^{-1-\mu}$, and the total number $K$ peaks around the 
value $S/\langle s \rangle$. Therefore, a tangible difference between the two models is
that the power-law regime has an exponent $1$ in the Sutton model and  
the size of the sectors rarely exceeds $\sqrt{S}$,
whereas the the distribution is a power-law with exponent $1+\mu \approx 2.35$ up to $S$ in our model
(with possibly a small hump for $s \sim S$, see Eq.~(\ref{pss})).
We hope that these falsifiable predictions of the two descriptions, as well 
as the quantitative description of the rescaled distribution of growth rates given 
above, 
will motivate further empirical and theoretical research, and help elucidate 
the `scaling puzzle' of company growth.

\section*{Note added:} While completing this work, X. Gabaix sent us a 
very interesting preprint 
where related arguments (although in details quite different from ours) are discussed. See: X. Gabaix, {\it Power-laws and the origin of the business cycle}, working paper, October 2002.

\section*{Acknowledgments} We want to thank X. Gabaix, M. Potters, J. Scheinkman and in particular J. Sutton for interesting discussions. J.P. B also wants to thank the organizers of the Bali conference on Econophysics that took place in August 2002 and that motivated this work.


\begin{thebibliography}{99}

\bibitem{Stanley} M. H. R. Stanley, L. A. N. Amaral, S. Buldyrev, S. Havlin, H. Leschorn, P. Maass, M.A. Salinger, H. E. Stanley, Nature, {\bf 319} 804 (1996)

\bibitem{Stanleybis} L. A. N. Amaral, S. V. Buldyrev, S. Havlin, H. Leschhorn, P. Maass,
M. A. Salinger, H. E. Stanley, and M. H. R. Stanley, ``Scaling Behavior
in Economics: I. Empirical Results for Company Growth'' J. Phys. I
France {\bf 7}, 621 (1997); S. V. Buldyrev, L. A. N. Amaral, S. Havlin, H. Leschhorn, P. Maass,
M. A. Salinger, H. E. Stanley, and M. H. R. Stanley, ``Scaling Behavior
in Economics: II. Modeling of Company Growth'' J. Phys. I France {\bf
7}, 635 (1997).
 

\bibitem{Stanley2} Y. Lee, L. A. N. Amaral, D. Canning, M. Meyer, H. E. Stanley, Phys. Rev. Lett. 
{\bf 81} 3275 (1998); D. Canning,  L. A. N. Amaral, Y. Lee, M. Meyer, H. E. Stanley, Economics Letters, {\bf 60} 335 (1998).

\bibitem{Sutton} G. Sutton, Physcia {\bf A} 312, 577 (2002).

\bibitem{partition} G. Andrews, {\it The Theory of Partitions}, Cambridge University Press (1998).

\bibitem{branchedpolymer} The $\sqrt{S}$ scaling found here for the typical size of sub-pieces allows to
shed light on a totally unrelated problem, that of `branched polymers' in high dimensions \cite{deG,Lubensky}. In the absence of steric constraints, the end-to-end distance of branched 
polymers grows like $N^{1/4}$, where $N$ is the total number of monomers. The scaling comes from the fact that a typical linear strand of the polymer contains $\sim N^{1/2}$ monomers, each of which behaving as 
a random walk in space. The analysis given here suggests that the number of independent linear strands 
(that plays the role of $K$) scales as $N^{1/2}\ln N$, with relative fluctuations that tend to zero, and that the size distribution of these strands should decay as $1/n$ for $n \ll N^{1/2}$.

\bibitem{deG} P. G. de Gennes, Biopolymers, {\bf 6} 715 (1968).

\bibitem{Lubensky} T. Lubensky, J. Isaacson, Phys. Rev. A {\bf 20}, 2130 (1979).

\bibitem{BM2000} J.-P. Bouchaud, M. M\'ezard,  Physica A {\bf 282}, 536 (2000).

\bibitem{Gabaix} X. Gabaix, Quarterly Journal of Economics, {\bf 114}, 739 (1999).

\bibitem{QF} For a short review, see: J.-P. Bouchaud, Quantitative Finance {\bf 1}, 105, (2000).

\bibitem{Bardou} F. Bardou, J.P. Bouchaud, A. Aspect, C. Cohen-Tannoudji, {\it L\'evy statistics and 
Laser cooling},  Cambridge University Press (2002).

\bibitem{PhysRep} J.P. Bouchaud, A. Georges, Phys. Rep. {\bf 195}, 127 (1991), ch. 3.

\bibitem{GL} C. Godreche, J.M. Luck, J. Stat. Phys. {\bf 104}, 489 (2001).

\bibitem{Derrida} B. Derrida, Physica D {\bf 107}, 186 (1997).

\bibitem{Axtell} R. Axtell, Science, {\bf 293}, 1818 (2001).

\end{thebibliography}
\end{document}